\documentclass[twocolumn]{aastex631}
\usepackage{silence}
\usepackage{graphicx}
\usepackage[caption=false]{subfig}
\usepackage{booktabs}
\usepackage{stmaryrd}
\usepackage{xcolor}
\usepackage{natbib}
\setcitestyle{authoryear,open={(},close={)}}

\usepackage{float}
\usepackage{color,soul}
\usepackage{wrapfig}
\usepackage{bm}
\usepackage{amsmath}

\usepackage{fancyhdr}
\usepackage{wasysym}

\begin{document}

\title{Modeling Galaxies in the Early Universe with Supernova Dust Attenuation}
\author[0000-0002-6149-8178]{Jed McKinney}
\altaffiliation{NASA Hubble Fellow}
\affiliation{Department of Astronomy, The University of Texas at Austin, Austin, TX, USA}

\author[0000-0003-3881-1397]{Olivia R. Cooper}\altaffiliation{NSF Graduate Research Fellow}
\affiliation{Department of Astronomy, The University of Texas at Austin, Austin, TX, USA}

\author[0000-0002-0930-6466]{Caitlin M. Casey}
\affiliation{Department of Physics, University of California Santa Barbara, Santa Barbara, CA, USA}

\author[0000-0002-8984-0465]{Julian B. Mu\~noz}
\affiliation{Department of Astronomy, The University of Texas at Austin, Austin, TX, USA}

\author[0000-0003-3596-8794]{Hollis Akins}\altaffiliation{NSF Graduate Research Fellow}
\affiliation{Department of Astronomy, The University of Texas at Austin, Austin, TX, USA}

\author[0000-0003-3216-7190]{Erini Lambrides}\altaffiliation{NPP Fellow}
\affiliation{NASA-Goddard Space Flight Center, Code 662, Greenbelt, MD, 20771, USA}

\author[0000-0002-7530-8857]{Arianna S. Long}
\affiliation{Department of Astronomy, University of Washington, Seattle, WA, USA}


\begin{abstract}
Supernova may be the dominant channel by which dust grains accumulate in galaxies during the first Gyr of cosmic time as formation channels important for lower redshift galaxies, e.g., AGB stars and grain growth, may not have had sufficient time to take over. Supernovae (SNe) produce fewer small grains, leading to a flatter attenuation law. 
In this work, we fit observations of 138 spectroscopically confirmed $z>6$ galaxies adopting standard spectral energy distribution modeling assumptions and compare standard attenuation law prescriptions to a flat attenuation law. 
Compared to SMC dust, flat attenuation close to what may be expected from dust produced in SNe yields up to $0.5$ mag higher $A_V$, and $0.4$ dex larger stellar masses. It also finds better fits to the rest-frame UV photometry with lower $\chi^2_{\rm UV}$, allowing the observed UV luminosities taken from the models to be fainter by $0.2$ dex on average. 
The systematically fainter observed UV luminosities for fixed observed photometry could help resolve current tension between the ionizing photon production implied by \textit{JWST} observations and the redshift evolution of the neutral hydrogen fraction. Given these systematic effects and the physical constraint of cosmic time itself, fairly flat attenuation laws that could represent the properties of dust grain produced by SNe should be a standard consideration in fitting to the spectral energy distributions of $z>6$ galaxies. 
\end{abstract}

\section{Introduction\label{sec:intro}}

The presence of dust grains in galaxies at very early times can significantly change inferred properties like total stellar mass and UV luminosity, key metrics for assessing the evolutionary pathways by which galaxy evolution begins. Historically, dust is thought to be sub-dominant in galaxies until $z\lesssim3-4$ \citep[e.g.,][]{Zavala2021} at which point galaxies have had enough time to grow dust reservoirs via Asymptotic Giant Branch (AGB) stars undergoing dust-enriched mass loss. However, direct observations of dust in emission exists up to $z=8.3$ \citep{Tamura2019} while indirect evidence from carbon enrichment, Balmer decrements, and UV continuum slopes extend to $z=12$ \citep{Bunker2023,Zavala2024,Langeroodi2024}. Recently, \cite{Schneider2024} review theoretical mechanisms that might produce dust at such early cosmic times, demonstrating that core collapse Supernovae (SNe) are the most likely grain-formation channel in the first Gyr given the relevant timescales and dust yields (see also \citealt{Todini2001,Nozawa2003}). By $z>6$ the age of the Universe is $<1$ Gyr which is prohibitively close to the $0.5$ Gyr needed for any low-mass ($\lesssim2M_\odot$) stars to evolve off the main-sequence. Other plausible mechanisms include dust formation in winds driven by active galactic nuclei \citep{Sarangi2019}, or ambient grain growth in the cold interstellar medium \citep{Draine1979,Draine2009,Popping2017} --- two channels with highly uncertain yields and end state grain properties, especially in exotic systems like the first galaxies. 

What is to be done considering the effects of dust on observations of very high-redshift galaxies? One option is to assume that the effects of dust attenuation are extremely low \citep{Ferrara2023,Ferrara2024lya,Ferrara2024alma,Ferrara2024bm,Ziparo2023,Fiore2023}. This is a reasonable assumption, but one that should be made explicitly clear in observational works handling properties highly sensitive to the presence of dust like total stellar mass. Even among the most obscured galaxies in the Universe, dust only constitutes $<1\%$ of the ISM by mass \citep{Scoville2017}. In other words, a little bit of dust goes a long way in shaping the observed, and intrinsic, properties of galaxies. Moreover, observations of very high-redshift galaxies typically do not probe the rest-frame optical so any dust constraint relies on extrapolating from the rest-frame UV which exacerbates any uncertainty arising from dust assumptions. 

Should our assumptions about $z>6-8$ dust change, photometry and spectra should be corrected using our knowledge of dust attenuation laws and dust formation pathways. This is now largely being done by applying well-understood attenuation laws tied to local Universe observations \citep[e.g.,][]{Topping2024}. This is well-motivated by the fact that SMC attenuation laws \citep{Gordon2003} have been shown to work well for low metallicity, high-redshift galaxies \citep{Reddy2015}, and also because Milky Way laws \citep{Cardelli1989} harbor the broad 2175\AA\ feature that has now been spectroscopically confirmed at $z=6.33$ \citep{Witstok2023}. None of these attenuation laws are explicitly representative of the dust grains expected to arise purely from SNe which could be producing most of the dust. Recent works such as \cite{Markov2023,Markov2024} and \cite{Fisher2025} are flexibly modeling the attenuation law shape directly, which is a promising approach for incorporating dust attenuation uncertainties into modeling results, and also for inferring trends in the underlying shape. There is growing evidence for shallower attenuation at $z>6$ which could arise from an abundance of SNe dust \citep[e.g.,][]{Ferrara2022,Markov2023,Markov2024,Fisher2025}.

The dust grain size distribution is one of the most important factors dictating the shape of an attenuation law \citep{Salim2020}, and grain size yields from SNe are not representative of those found in the SMC, or the Milky Way on-average. SNe dust is generally characterized by a deficit in the small grains leading to gray UV attenuation \citep[e.g.,][]{Nozawa2003,Hirashita2005}. 
In this letter, we argue that the first approach to accounting for dust in galaxies at high-redshift ($z\gtrsim6-8$) should be to assume a fairly flat attenuation law as might be expected to arise from dust produced in SNe alone. This assumption has precedent among highly obscured $z\sim1$ galaxies \citep{Kawara2011,Shimizu2011}, quasars at the tail end of the epoch of reionization \citep{Hirashita2005,Bianchi2007}, and is consistent with the findings of \cite{Markov2023,Markov2024}, \cite{Langeroodi2024}, and \cite{Fisher2025} at $z>6$. In Section \ref{sec:attenuation} we compare attenuation laws and discuss the SNe law that we adopt in this analysis. Section \ref{sec:obs} presents the sample of galaxies that we fit under the SNe dust assumption with spectral energy distribution modeling as described in Section \ref{sec:sedfits}. We summarize our results in Section \ref{sec:results} and discuss their implications in Section \ref{sec:discussion}. Throughout this work we assume a $\Lambda$CDM cosmology with $H_0=70$\,km\,s$^{-1}$\,Mpc$^{-1}$, $\Omega_m=0.3$, $\Omega_\Lambda=0.7$, a Kroupa initial mass function (IMF, \citealt{Kroupa2001}), and the AB magnitude system \citep{Oke1974}. 

\begin{figure}
    \centering
    \includegraphics[width=0.48\textwidth]{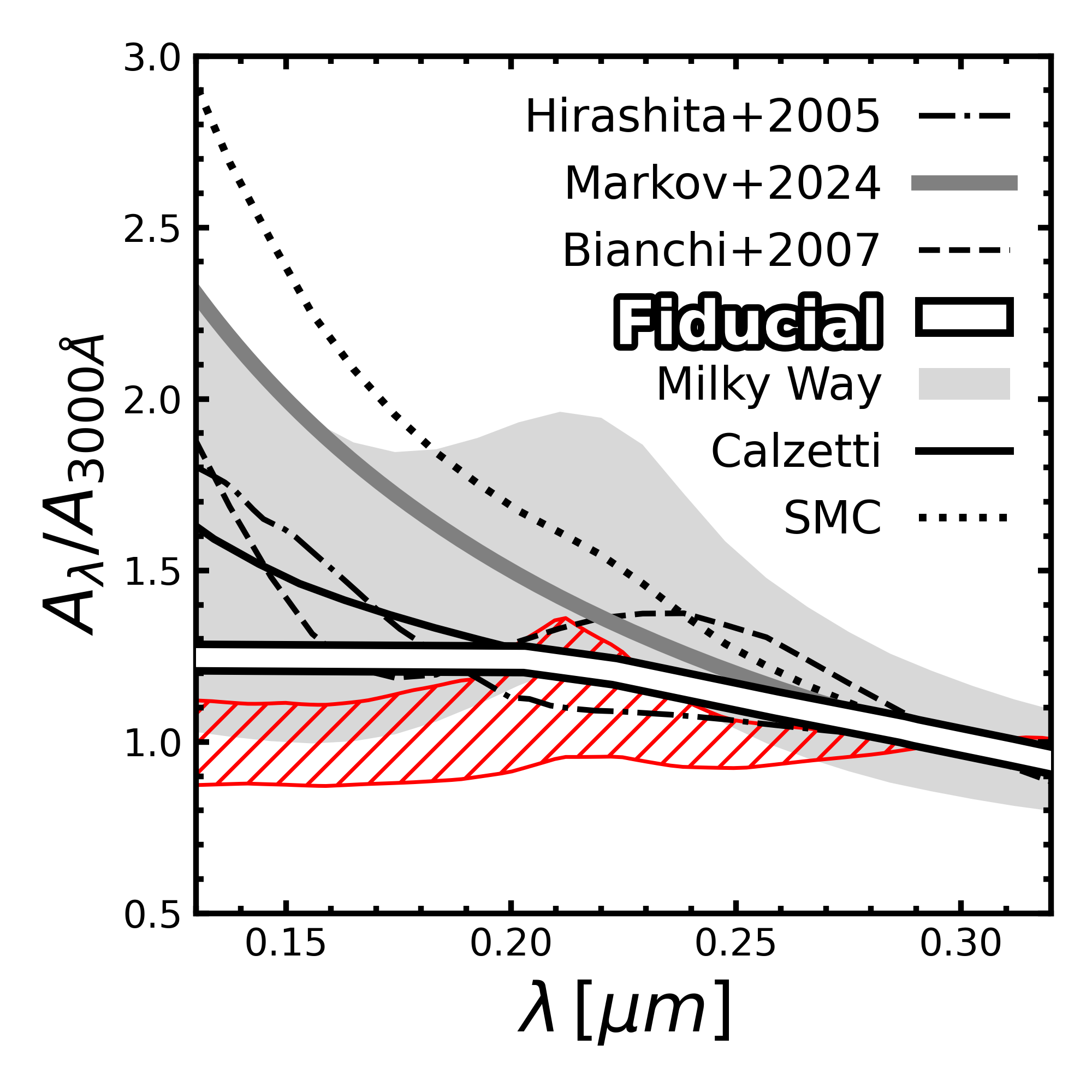}
    \caption{UV attenuation laws normalized at 3000\AA\ for various dust models and observations. Milky Way attenuation laws span the range shaded in gray, and the SMC law, often taken as standard for very high-redshift galaxies, is shown with the dotted line. The thick gray line shows the inferred attenuation law from \cite{Markov2024} who fit to JWST spectra of galaxies with $6.3<z<11.5$. In solid black we show the empirical attenuation law measured by \cite{Gao2020}, which we adopt as our fiducial attenuation law. This empirical attenuation law is comparable to the theoretical extinction models of SNe dust from \cite{Hirashita2005} and \cite{Bianchi2007}, although the models have steeper UV slopes. The \cite{Gao2020} attenuation law has a comparable UV slope to the attenuation laws found for simulated $z>8$ galaxies from \cite{Narayanan2025} (red hatched), which are almost completely flat throughout the UV and optical. 
    }
    \label{fig:av}
\end{figure}

\section{The Supernova Dust Attenuation Law\label{sec:attenuation}}

We begin by describing the attenuation law $A_\lambda$ properties of SNe dust which has been constrained observationally and also modeled theoretically. \cite{Hirashita2005} investigate the extinction curves arising from Pop III into either Sne Type II or pair instability SNe (PISNe) both based on the Pop III models and elemental yields of \cite{Nozawa2003}. For larger SNe II progenitor masses ($\approx20M_\odot$) and PISNe both with well-mixed helium cores, the Pop III models predict very steep extinction in the UV tied to a high abundance small grains. Less massive SNe II progenitors ($\approx13M_\odot$) from Pop III stars and PISNe that do not have mixed helium cores, preserving the layered elemental structure, have much flatter extinction curves in the UV arising from the absence of very small grains. The smallest grains experience the most destruction by thermal sputtering over the course of the reverse shock \citep[e.g.,][]{Nath2008}. Indeed the evolutionary phases of SNe and particularly the reverse shock have a profound impact on the grain properties and therefore their corresponding attenuation law (see \citealt{Micelotta2018} for a thorough review). To characterize the processed properties of SNe dust, \cite{Bianchi2007} evolve a dust model of core collapse SNe ejecta through the reverse shock. The extinction models of \cite{Bianchi2007} show a mild 2175\AA\ bump and is more gray (i.e., flat) in the UV than MW and SMC-like attenuation laws as shown on Figure \ref{fig:av}, although scattering effects could bring these extinction laws into further agreement with MW and SMC-like attenuation.  


For the purposes of testing the impact of SNe dust attenuation laws on the \textit{JWST} observations of $z>6$ galaxies we adopt the empirical attenuation law derived in in \cite{Gao2020} from observations of the red type Ia SN 2012cu in NGC 4772, a low-luminosity dwarf Seyfert galaxy. \cite{Gao2020} infer a gray attenuation law coupled to a dust grain size distributed characterized by an absence of very small grains. Type Ia SNe are not expected to produce the dust in very early galaxies, and the measured $A_\lambda$ in \cite{Gao2020} is impacted by dust in the interstellar medium that may or may not arise purely from SNe. However, the underlying dust grain size distribution that \cite{Gao2020} recover is largely devoid of very small grains with a mean grain radius of $\bar a=0.16\,\mu m$ that is representative of the grain size distribution that might arise from SNe dust at early cosmic times. Indeed, \cite{Nozawa2007} study the propagation of shocked Type II SNe dust grains into the ISM and find that grains with $a\lesssim0.05\,\mu m$ are completely destroyed, and also that only grains with $a\gtrsim0.2\,\mu m$ survive transport into the ISM and do so without changing their size. This leads to dust grain size distributions weighted significantly towards the larger grain population, in line with what \cite{Gao2020} infer from SN 2012cu in NGC 4772.

As shown on Figure \ref{fig:av} the shape of $A_\lambda$ from \cite{Gao2020} 
traces the median of the SNe dust extinction models from \cite{Hirashita2005} and \cite{Bianchi2007} at wavelengths $>1500$\AA. At shorter wavelengths the models exhibit steeper UV slopes, hinting at a residual small grain population, and are similar in shape to the SMC law but with $>2$ mag less extinction for fixed $A_{3000\dot{\rm A}}$ or $A_V$ relative to SMC $A_\lambda$. 
To explore the strict assumption on a very large dust grain size distribution of SNe origin that is attenuating stellar light, we adopt the empirical $A_\lambda$ from \cite{Gao2020} as our fiducial attenuation law. In later sections we discuss the implications of instead using the theoretical attenuation law from \cite{Hirashita2005}.

\cite{Markov2023,Markov2024} and \cite{Fisher2025} take a different approach by modeling the attenuation law directly whereby the parameters controlling its slope at different wavelengths as well as the 2175\AA\ bump strength vary freely during fits. The mean attenuation law \cite{Markov2024} derive for galaxies with $6.3<z_{\rm spec}<11.5$ is shown on Figure \ref{fig:av} and exhibits greater attenuation in the UV compared to a pure SNe dust law, but less attenuation relative to the SMC law. \cite{Markov2024} and \cite{Fisher2025} both note that this likely arises from grain reprocessing, possibly driven by SNe shocks preferentially destroying the small grains. Our approach in this work is complimentary to that of \cite{Markov2023,Markov2024} and \cite{Fisher2025}. We test the limiting assumption of a flat attenuation law which removes degeneracies with other stellar population synthesis properties that may otherwise change the UV slope. Both approaches are needed to test the nature of dust extinction and attenuation at very early cosmic times.

\begin{figure}
    \centering
    \includegraphics[width=0.48\textwidth]{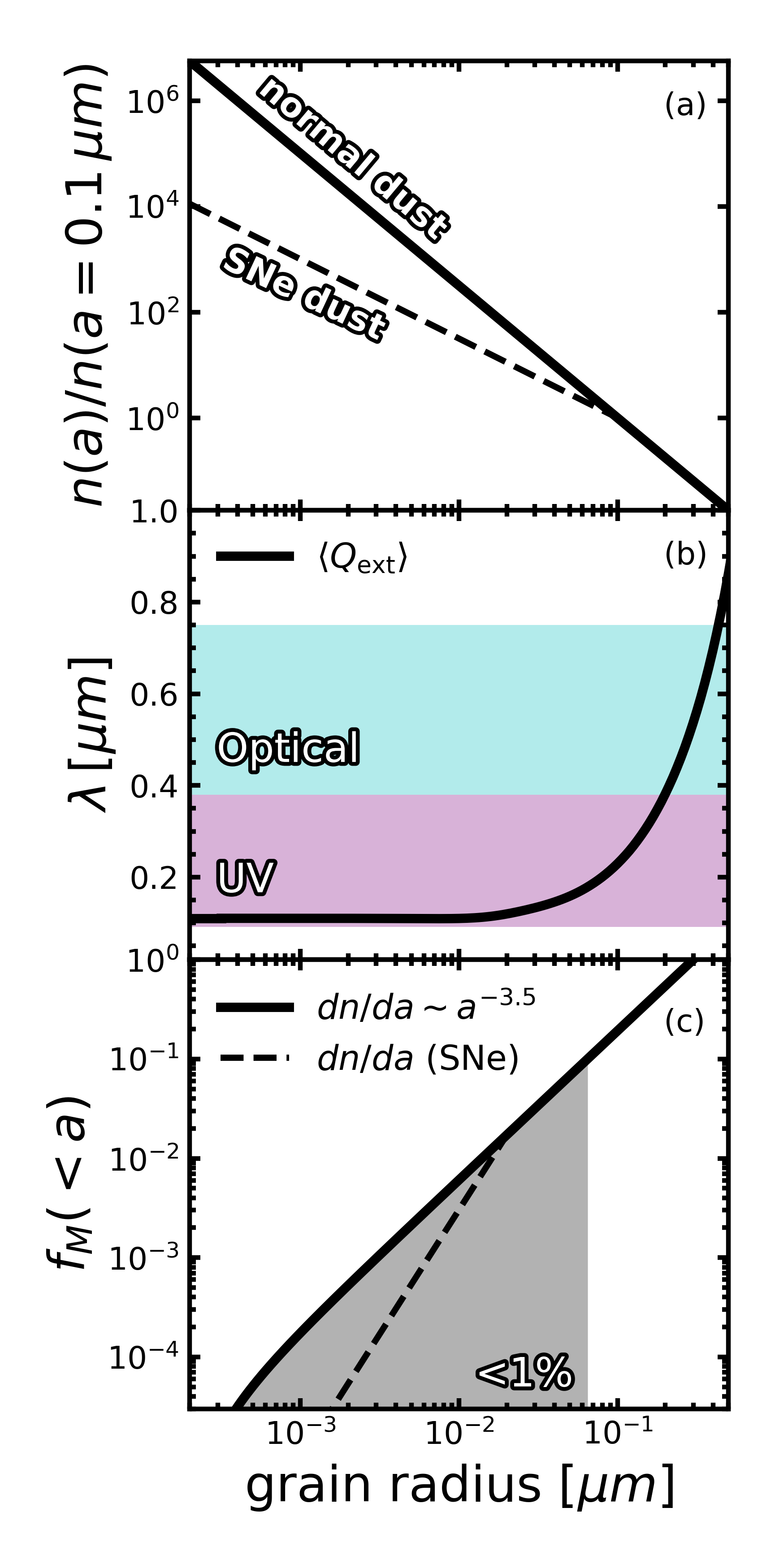}
    \caption{(\textit{a}) Grain size distribution normalized at $0.1\,\mu$m for standard dust (solid, \citealt{Mathis1977}) and SNe dust (dashed, \citealt{Nozawa2003}) with a break below $\approx0.02\,\mu$m. (\textit{b}) The rest-frame wavelength corresponding to the average extinction efficiency $Q_{\rm ext}$ for a grain of size $a$, based on the \texttt{Astrodust} model \citep{Draine2021,Hensley2023}. Below $a\approx0.016\,\mu$m $Q_{\rm ext}$ converges to the Lyman limit which is also the wavelength limit of \texttt{Astrodust}. Grains with $a\lesssim0.2\,\mu$m contribute to UV attenuation, but the total UV attenuation is dominated by very small grains with $a\lesssim0.01\,\mu$m that are $\sim10^4\times$ more numerous because of the steep power-law grain size distribution. (\textit{c}) The fraction of mass in dust grains with size $<a$. In all cases 90\% of the total dust mass is accounted for by grains with $a>0.1\,\mu$m that contribute negligibly to the total UV attenuation.}
    \label{fig:fm}
\end{figure}

\subsection{The impact of attenuation laws on the inferred total dust mass\label{sec:sne:dustmass}}
Dust grains responsible for attenuating UV light comprise a negligible fraction of the total dust mass. Therefore, the same quantity of dust can shape the observed UV spectrum of galaxies in very different ways depending on the underlying grain size distribution. 

The variations between steep or shallow UV attenuation arise from small dust grains with grain radii $a<0.1\,\mu m$. Figure \ref{fig:fm} (\textit{a}) shows the grain size distribution for standard and SNe dust exhibiting a deficit of small grains below $a\approx0.02\,\mu m$ \citep{Nozawa2003}. Grains with sizes below $a\approx0.01\,\mu m$ are $\sim10^4\times$ more numerous than grains with $a=0.1\,\mu m$. Figure \ref{fig:fm} (\textit{b}) shows the wavelength corresponding to the average extinction efficiency $Q_{\rm ext}$\footnote{For this calculation we adopt the tabulated extinction efficiencies of the \texttt{Astrodust} model as described in \cite{Hensley2023} and \cite{Draine2021}.} for a grain of size $a$. Grains with $a\lesssim0.1\,\mu m$ all contribute to UV attenuation but the smallest grains ($a\lesssim0.01\,\mu m$) are dominant because they preferentially absorb at shorter wavelengths and are far more numerous. Assuming the standard power-law distribution in grain sizes $dn/da\sim a^{-3.5}$ \citep{Mathis1977}, the total mass of dust grains $M_d$ with sizes between $a_-$ and $a_+$ is
\begin{equation}
    \begin{split}
    M_d\sim\int_{a_-}^{a_+} m(a)n(a)\, da &\sim \int_{a_-}^{a_+} \frac{4}{3}\pi\bar\delta_g a^3 a^{-2.5}\, da
    \\ 
    &\sim \int_{a_-}^{a_+} \sqrt{a}\,da \propto \big[a^{3/2}\big]^{a_+}_{a_-}
    \end{split}
\end{equation}
where $\bar\delta_g$ is the mean grain density. Thus the fractional mass of grains of size $<a$ in the truncated range of $[a_-,a_+]=[2\times10^{-4}\,\mu m, 0.3\,\mu m]$ is
\begin{equation}
    f_M(<a)=\frac{a^{3/2}-a_-^{3/2}}{a_+^{3/2}-a_-^{3/2}}
\end{equation}
as shown in Figure \ref{fig:fm}\textit{c}. Based on the calculations of \cite{Nozawa2003}, the size distribution of SNe dust has a characteristic break below $a_{crit}\approx0.02\,\mu m$ where $dn/da \sim a^{-2.5}$, reflective of the shift towards larger grain sizes on-average. In this regime we have $M_d(a)\propto a^{2.5}$ following the same approach. 

Figure \ref{fig:fm}\textit{c} shows the fractional mass in grains of size $a$ between $2\times10^{-4}\,\mu m$ and $0.3\,\mu m$. In both the standard power-law size distribution and for a SNe break below $0.02\,\mu m$, grains of size $a>0.1\,\mu m$ contribute $>90\%$ to the total dust mass. The grains most responsible for UV absorption have $a\approx0.01\,\mu m$ or smaller and account for $<1\%$ of the total dust mass. As a result, the use of SNe attenuation laws will not lead to an observed shift on $M_d$, and total dust masses will agree with what is otherwise allowed by other attenuation laws.

\begin{figure*}[t]
    \includegraphics[width=\textwidth]{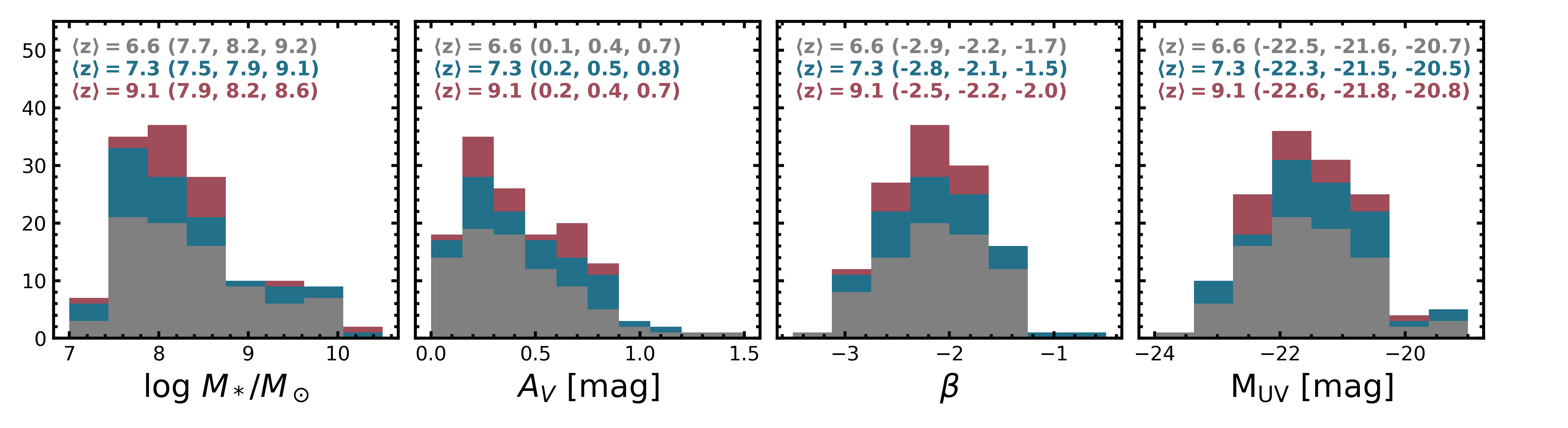}
    \caption{Stacked histograms of stellar mass, $A_V$, UV slope $\beta$, and total UV magnitude ${\rm M_{UV}}$ for our sample sub-divided into three redshift bins: $z<7$ with mean $z=6.6$ (black), $7<z<8$ with mean $z=7.3$ (blue), and $z>8$ with mean $z=9.1$ (red). We report the 16th, 50th, and 84th percentiles in each redshift bin at the top of each panel.}
    \label{fig:hists}
\end{figure*}

\section{High-Redshift Galaxy Observations\label{sec:obs}} 
Prior to \textit{JWST} the number of spectroscopically-confirmed galaxies having redshifts greater than 7 was fewer than 20 \citep[e.g.,][]{Ouchi2020}. Deep \textit{JWST} imaging surveys and their spectroscopic follow-up programs have now revealed hundreds of galaxies with confirmed redshifts up to $z\sim14$ (see \citealt{Stark2025} for a review), thus providing excellent statistical samples to begin critically revising our assumptions about galaxies in the very early Universe.  

Given that dust attenuation laws vary more significantly in the UV, the assumption of flat attenuation law as a model for SNe dust is likely to have significant impact on the interpretation of UV continuum slopes ($\beta$), and total UV magnitudes ($M_{\rm UV}$). The relatively shallower SNe dust attenuation laws may also allow greater values of $A_V$ while preserving the observed UV spectral shape relative to SMC laws, which can impact the inferred total stellar mass. To build a robust sample for testing the impacts of SNe dust on these properties we select galaxies from the \textit{JWST} Advanced Deep Extragalactic Survey (JADES, \citealt{Eisenstein2023}, \dataset[https://doi.org/10.17909/8tdj-8n28]{https://doi.org/10.17909/8tdj-8n28}). We take photometry in NIRCam medium and broad bands from \cite{Eisenstein2023b} based on the maps of \cite{Rieke2023}. We also adopt spectroscopic redshifts based on NIRSpec/MSA spectroscopy from JADES \citep{Bunker2024,DEugenio2024}, the complementary NIRCam/grism spectra from the First Reionization Epoch Spectroscopically Complete Observations survey (FRESCO, \citealt{Oesch2023}), and further NIRCam imaging from the JWST Extragalactic Medium-band Survey (JEMS, \citealt{Williams2023}, \dataset[https://doi.org/10.17909/fsc4-dt61]{10.17909/fsc4-dt61}). These data cover the GOODS-North and GOODS-South extragalactic legacy fields with ancillary multi-wavelength imaging from ground and space-based telescopes like \textit{HST}. These galaxies are broadly representative of the average galaxies \--- in mass, UV color, and luminosity \--- of what has been observed to date with \textit{JWST} with no reason to suspect a bias in dust characteristics.

From the JADES public catalog \citep{Eisenstein2023b} we select all objects with spectroscopic redshifts $z\geq6$ derived from at least one emission line in the medium-resolution grating, and/or two or more MSA prism lines. This yields a high-fidelity sample of 138 galaxies, with 82 ($60\%$) having $6<z_{\rm spec}<7$, 35 ($25\%$) having $7<z_{\rm spec}<8$, and 19 ($14\%$) having $8<z_{\rm spec}<10$. Two sources have $z>10$. We adopt Kron aperture photometry PSF-matched to the NIRCam/F444W band following the convention of \cite{Topping2024} who also fit and analyze the UV slopes of JADES galaxies, but under the assumption of an SMC-like attenuation law. In this manner we collect PSF-matched photometry in the following NIRCam wide bands: F090W, F115W, F150W, F200W, F277W, F356W, and F444W as well as the medium band filters F182M, F210M, F335M, F410M, F430M, F460M, and F480M. We supplement these with PSF-matched photometry from \textit{HST} also available from the JADES catalog in F435W, F606W, F814W, F850LP, F105W, F125W, F140W, and F160W. 

We re-measure empirical UV slopes ($\beta$) following \cite{Topping2024} by fitting $f_\lambda\sim\lambda^{-\beta}$ to the F115W, F150W, and F200W filters for $5.4<z_{\rm spec}<7.4$, F150W, F200W, and F277W for $7.4<z_{\rm spec}<9.8$ and F200W, F277W, and F356W for $z_{\rm spec}>9.8$. To perform this fit we use the non-linear least-squares minimization algorithm as implemented in \texttt{lmfit} \citep{lmfit}, which propagates the photometric uncertainties into the final errors on $\beta$. Our measured UV slopes are fully consistent with \cite{Topping2024} which is expected given the sample overlap.

\section{SED Fitting with Supernova dust\label{sec:sedfits}}

We fit the observed spectral energy distributions (SEDs) of our sample using \texttt{BAGPIPES} \citep{Carnall2018} powered by \texttt{Nautilus} \citep{nautilus}, the latter of which implements a neural network approach to Bayesian posterior estimation. For a full description of \texttt{BAGPIPES} see \cite{Carnall2018}. We adopt the standard assumptions of \texttt{BAGPIPES} as described therein. For each target we perform two runs with otherwise identical priors (see below) but with different attenuation laws. For the SNe dust model we adopt the empirical attenuation law from \cite{Gao2020} as discussed in Section \ref{sec:attenuation}. For the SMC model we adopt the standard SMC attenuation law \citep{Gordon2003} as implemented in the default \texttt{BAGPIPES} configuration \citep{Carnall2018}. 

\begin{figure*}
    \centering
    \gridline{\fig{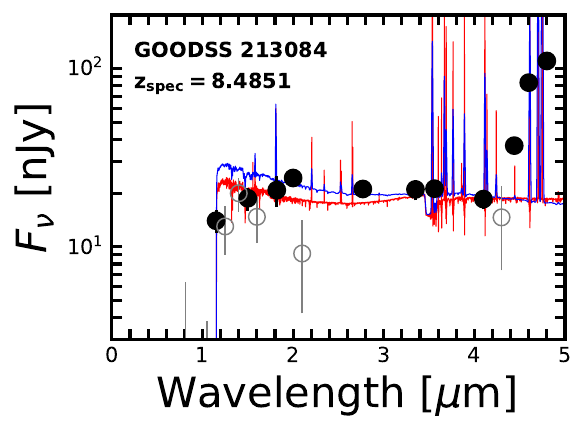}{0.32\textwidth}{}
              \fig{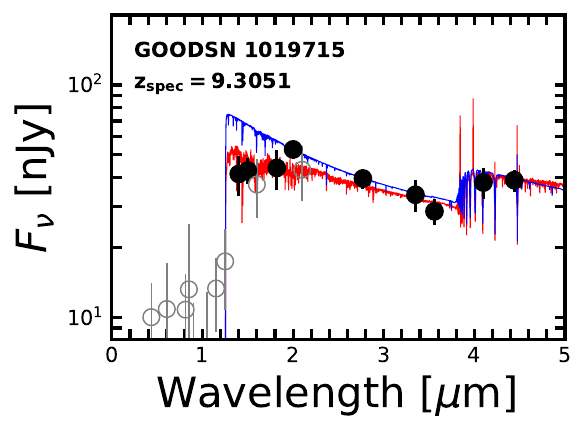}{0.32\textwidth}{}
              \fig{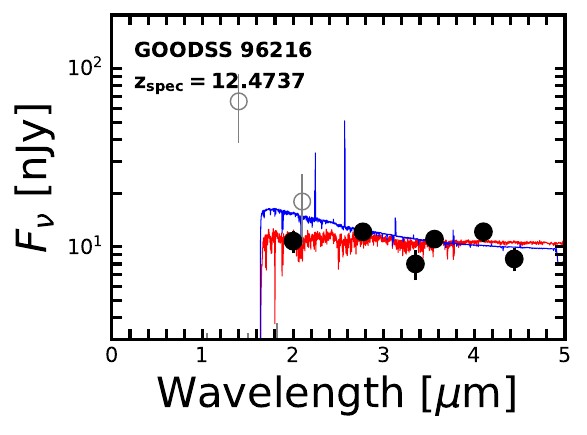}{0.32\textwidth}{}}
    \vspace{-20pt}
    \caption{Example SED fits to the \textit{JWST}/NIRCam photometry of spectroscopically confirmed $z>8$ galaxies from \cite{Eisenstein2023b,Bunker2024} and \cite{DEugenio2024}. Photometry shown with filled (open) circles have SNR$\,>5$ (SNR$\,<5$). Red and blue SEDs show the maximum likelihood models when assuming flat attenuation for SNe dust and SMC dust attenuation respectively.}
    \label{fig:seds}
\end{figure*}

When fitting each target we fix the model redshift to the observed spectroscopic redshift. We include a nebular component with a radiation field density allowed to vary between $\log\,U\in(-4,0)$ to cover the range of values observed in galaxies at $z>6$ \citep[e.g.,][]{Tang2023}. We set an upper bound on the total stellar mass of $10^{11}\,M_\odot$ which in practice is never reached, and we let the metallicity vary between $10^{-4}Z_\odot$ and $Z_\odot$. We let $A_V$ vary between $0-2$. For each free parameter we assume uniform priors which are logarithmic for stellar mass, metallicity and nebular radiation density and are linear otherwise. 

The choice of star-formation history can impact the final derived parameters. In this work we adopt the non-parametric model of \cite{Leja2019} which enforces a continuity prior between prescribed bins of variable SFR. We follow the parameter choices of \cite{Markov2023,Markov2024} who fit similar galaxies as those in our sample but with the intent of inferring the attenuation law as opposed to assuming its shape. We adopt $N=7$ logarithmically spaced bins between the age of the Universe at each galaxy's redshift out to a look-back time of $z=20$. The parameter posteriors are generally insensitive to the number of bins for $N=4-14$ \citep{Leja2019}. We measure the posterior distribution in observed $\rm{M_{UV}}$ by integrating the posterior SEDs convolved with a top hat filter between 1450\AA\ and 1550\AA. 

We are most interested in the impact of SNe dust on the UV slope, the total UV luminosity, $A_V$, and the total stellar mass. These are well-constrained by the JADES filters that can recover the underlying continuum shape as well as the contamination in broad bands by strong emission lines thanks to the medium band coverage and sensitivity. Indeed \cite{Yanagisawa2024} find derived properties from NIRCam photometry and NIRSpec to be consistent within the errors. Thus fitting only the photometry to back out the aforementioned quantities is well-motivated. \texttt{BAGPIPES} is capable of fitting to the available spectroscopy \citep{Carnall2019b} and implements a wavelength-dependent slit-loss correction polynomial that in principle is a function of the target's morphology and the slit placement. This correction term can contribute to the models' rest-frame continuum UV shape, which is also impacted by the star-formation history and dust attenuation. For this reason we choose to fit only the photometry, and leave fitting the joint photometry and spectra to future work. Example fits to the data for galaxies at $z\sim8-12$ are shown in Figure \ref{fig:seds}.

\begin{figure}
    \centering
    \includegraphics[width=0.5\textwidth]{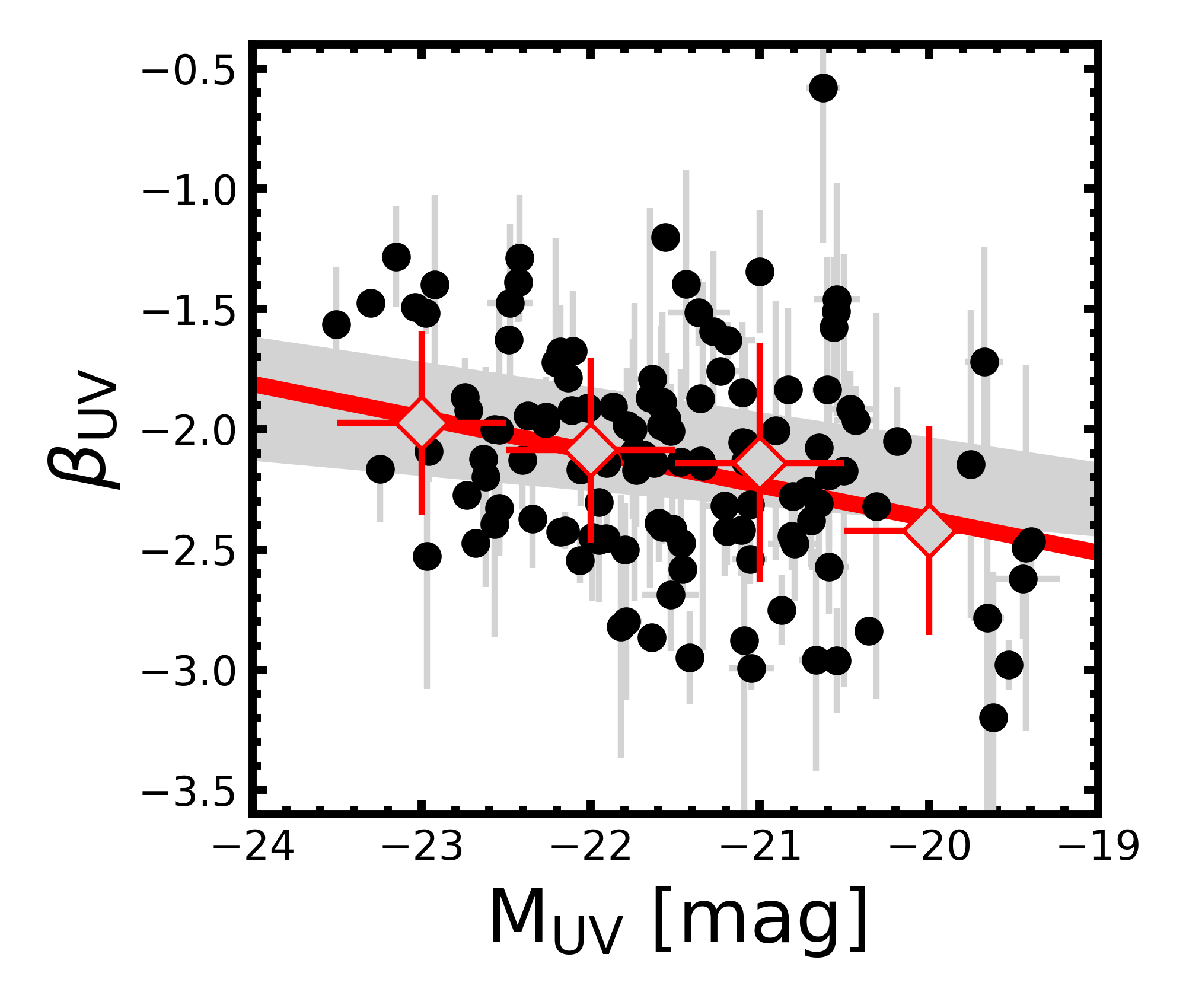}
    \caption{Empirical UV slopes $\beta$ vs.~the observed UV magnitudes taken from the best-fit models (black circles). The red diamonds show the weighted-average of $\beta$ in bins of $\Delta{\rm M_{UV}}=2$. The grey region encases the trends from \cite{Topping2024} for sources with $z\sim6-9$. The red line shows a linear fit to the binned average. Our results are generally consistent with the $\beta-{\rm M_{UV}}$ trends of prior studies even with different dust prescriptions.}
    \label{fig:bmuv}
\end{figure}

\begin{figure}
    \centering
    \includegraphics[width=0.47\textwidth]{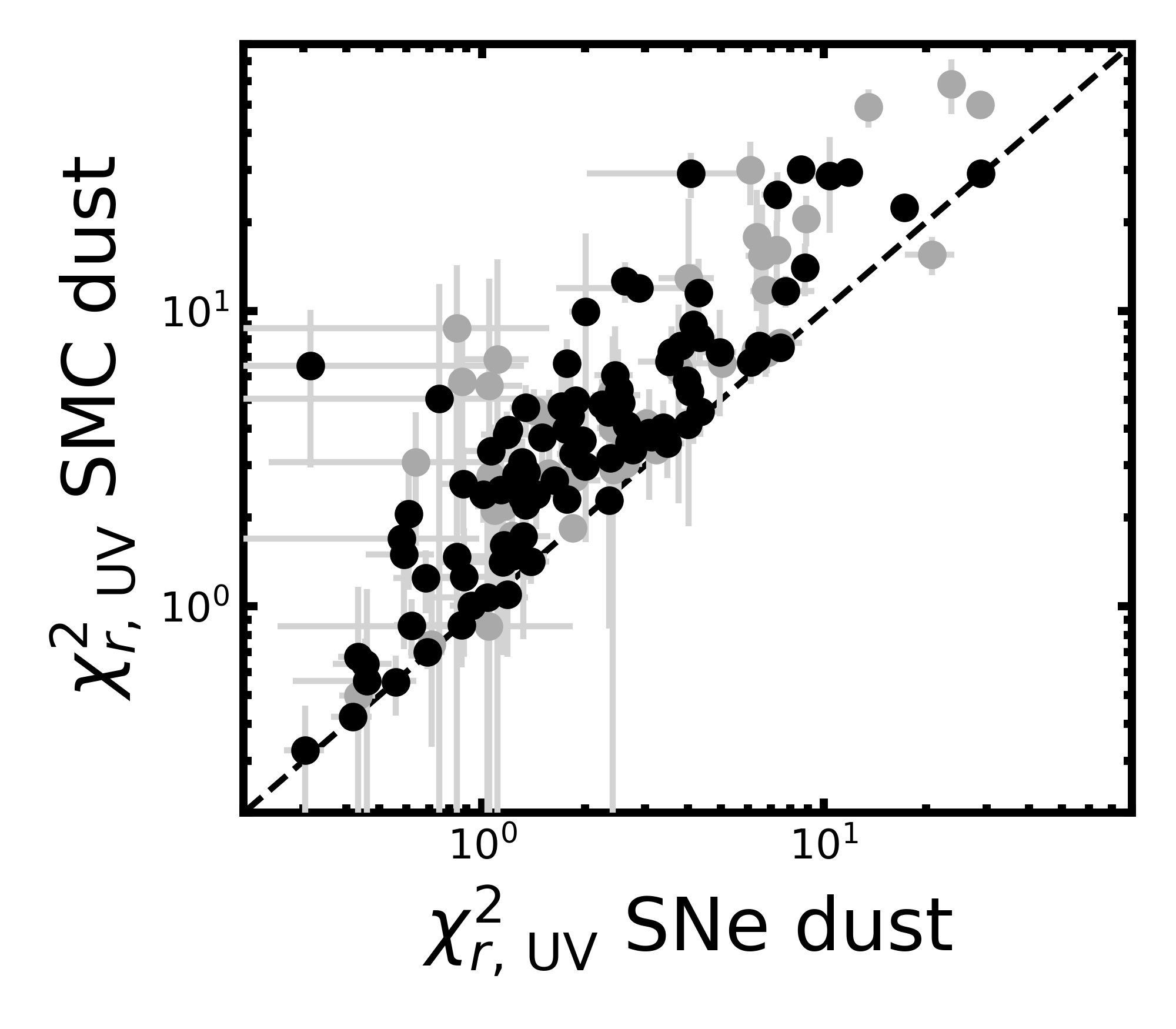}
    \caption{Reduced-$\chi^2$ from models fits using SNe vs.~SMC dust attenuation calculated in the rest-frame UV. The 1-to-1 relation is shown with the dashed black line, and black (gray) circles have consistent (inconsistent) star-formation histories as defined in Section \ref{sec:res:smc}. Fairly flat dust attenuation laws find better fits to the rest-frame UV data of $z=6-12$ galaxies.}
    \label{fig:chi2uv}
\end{figure}

\section{Results\label{sec:results}}

\subsection{Consistency with prior works}
Figure \ref{fig:hists} shows histograms for quantities derived from our observations and spectral energy distribution modeling. We measure stellar masses between $10^7{M_\odot}$ and $10^{10}{M_\odot}$ with $A_V\in(0,1.5)$ but most of the sample exhibit $A_V<1$ which is within the range of prior works fitting for the attenuation in $z>6$ galaxies using \textit{JWST}/NIRCam and/or \textit{JWST}/NIRSpec data \citep{Markov2023,Markov2024,Langeroodi2024,Fisher2025}. We measure UV slopes between $-3<\beta<-1$, typical of other studies \citep{Topping2024,Austin2024,Cullen2024,Morales2024}, which is expected as these are measured direct from the data and do not depend on our SNe dust assumptions. We infer observed ${\rm M_{UV}}\in(-24,-19)$ which is consistent with the range in UV luminosities reported in prior works that in some cases adopt different spectral energy distribution model assumptions \citep[e.g.,][]{Bunker2023,Topping2024}. As shown Figure \ref{fig:bmuv},  we are able to recover the anti-correlation between UV slope and observed UV luminosity found in prior works \citep[e.g.,][]{Topping2024} that do not use SNe dust. 

\subsection{Comparison with SMC-like attenuation\label{sec:res:smc}}
\begin{figure}[t]
    \centering
    \includegraphics[width=\linewidth]{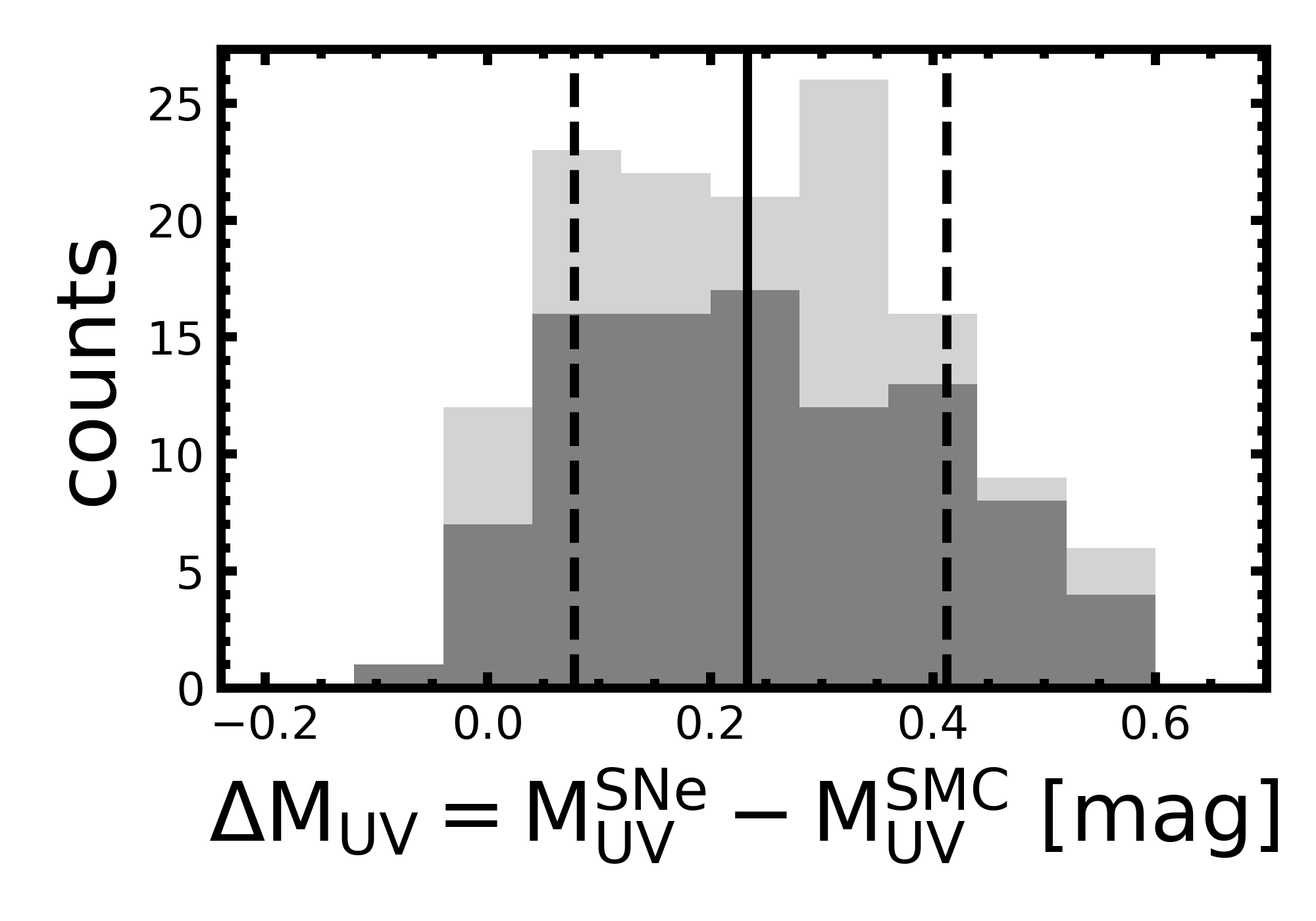}
    \caption{Stacked histogram showing the change in ${\rm M_{UV}}$ measured from the model SEDs assuming SNe vs.~SMC dust attenuation. The dark (light) gray histogram corresponds to sources with convergent (divergent) star-formation histories between the two fitting assumptions. Fairly flat attenuation leads to fainter observed ${\rm M_{UV}}$ by $0.23^{+0.18}_{-0.16}$ mag.}
    \label{fig:dmuv}
\end{figure}

As described in Section \ref{sec:sedfits} we perform another set of \texttt{BAGPIPES} fits to the data with identical priors but adopting an SMC attenuation law. Formally the two fits yield similar likelihood distributions per source and over the full sample. This is largely a product of the models' success in reproducing the redder NIRCam bands that have high signal-to-noise ratios, especially when strong emission lines dominate their flux densities. 
Figure \ref{fig:chi2uv} shows the reduced-$\chi^2$ calculated only with photometry covering the rest-frame UV. Flat attenuation models possibly representative of SNe dust attenuation yield better fits to the UV than those assuming SMC-like dust attenuation. As a result, the observed UV luminosities measured from these fits are fainter by $0.23^{+0.18}_{-0.16}$ mag as shown in Figure \ref{fig:dmuv}.

\begin{figure}[t]
    \centering
    \includegraphics[width=\linewidth]{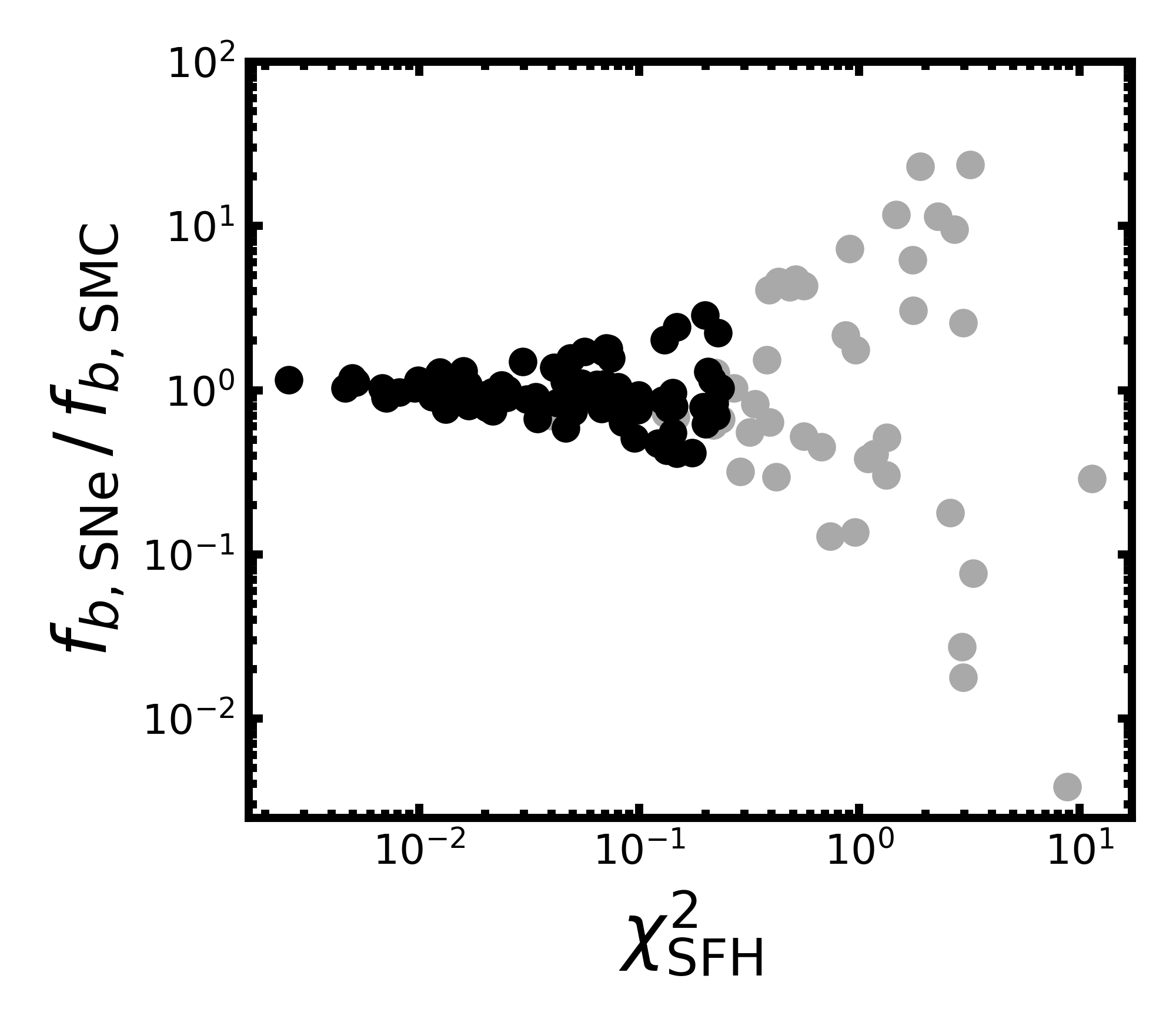}
    \caption{Star-formation history consistency between fits with SNe and SMC dust, using the $10$ Myr burst fraction $f_b$ and $\chi^2_{\rm SFH}$ as defined in Equation \ref{eq:chi2}. We consider star-formation histories between these two runs to be consistent (black) when the burst fractions agree within 20\% or $\chi^2_{\rm SFH}<0.25$. 96/138 (70\%) fits are consistent, and the remaining 30\% prefer different star-formation histories when changing just the attenuation law.}
    \label{fig:sfh}
\end{figure}

When comparing the two model runs we want to control for variations in the preferred star-formation histories. To quantify this we calculate a $\chi^2$-like statistic from the star-formation histories. In \texttt{BAGPIPES}' non-parametric implementation each age bin is assigned a $\Delta\log{\rm SFR}_n\equiv\log{\rm SFR_n/SFR_{n+1}}$ which is fit freely under a continuity prior \citep{Leja2019}. From these quantities output for both the SNe dust and SMC dust fits we calculate 
\begin{equation}
    \chi^2_{\rm SFH}\equiv\sum_n \big[\Delta\log{\rm SFR}_{n,\,\rm SNe}-\Delta\log{\rm SFR}_{n,\,\rm SMC}\big]^2
    \label{eq:chi2}
\end{equation}
which captures the total difference between the two star-formation histories. The 16th, 50th and 84th percentiles on the square of $\chi^2_{\rm SFH}$ are 0.1, 0.3, and 0.9. We also calculate a burst fraction $f_b$ defined as the fraction of star-formation within the most recent (10 Myr) age bin relative to the integrated star-formation history. Figure \ref{fig:sfh} compares these metrics across the sample. We define consistency as having $\chi^2_{\rm SFH}\leq0.25$ dex or $f_b$ within 20\% of one-another. Approximately $70\%$ of the fits have consistent star-formation histories between the SNe and SMC attenuation runs, and we find no significant bias against the remaining 30\% in $\beta$, $\rm{M_{UV}}$, $A_V$ or stellar mass.

\begin{figure}
    \centering
    \includegraphics[width=0.5\textwidth]{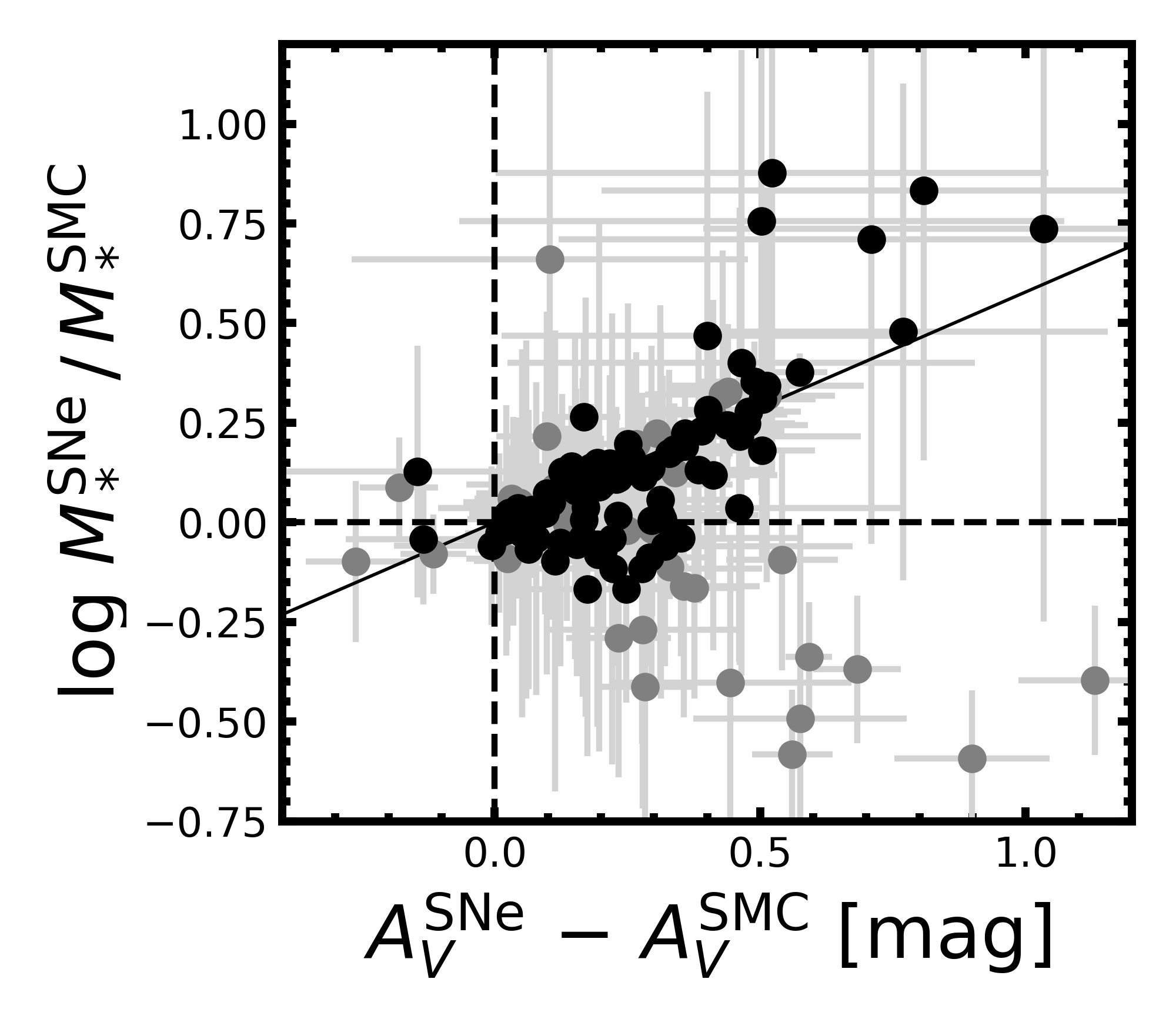}
    \caption{Change in stellar mass and $A_V$ when assuming SNe vs.~SMC dust attenuation. Solid black circles show fits that converge to similar star-formation histories under both attenuation law assumptions, whereas gray circles have divergent star-formation histories.
    The solid black line shows the best fit $\Delta\log M_*=0.6\Delta A_V$ to fits with convergent star-formation histories. We find that modeling SNe dust with flat attenuation leads to higher stellar mass estimates by up to $\sim0.8$ dex and $A_V$ by up to $\sim1$ mag.}
    \label{fig:smcsne}
\end{figure}

Figure \ref{fig:smcsne} shows the relative change in $A_V$ and stellar mass when assuming fairly flat dust vs.~SMC dust attenuation. Relative to the SMC law, fairly flat dust attenuation yields higher values for $A_V$ and for most of the sample correspondingly larger stellar masses. The 16th, 50th, and 84th percentiles for $\Delta A_V\equiv A_V^{\rm SNe}-A_V^{\rm SMC}$ and $\Delta\log M_*\equiv\log\,M_*^{\rm SNe}/M_*^{\rm SMC}$ are $(0.07,0.23,0.46)$ and $(-0.1, 0.1, 0.2)$ respectively. For the subset with consistent star-formation histories we find 16th, 50th, and 84th percentiles for $\Delta\log M_*$ of $(0,0.1,0.3)$ dex and $(0.1,0.2,0.5)$ mag for $\Delta A_V$. The fits that yield consistent star-formation histories follow a relation characterized by $\Delta\log M_*\,/\,\Delta A_V=0.6$.  On-average, the stellar masses increase by $\sim0.1$ dex when modeling SNe dust as a fairly flat attenuation law, comparable to the systematics found when varying more standard attenuation laws in lower redshift galaxies \citep{Reddy2015,Salim2016}, as well as parametric attenuation laws out to $z\sim10$ \citep{Markov2023,Markov2024,Fisher2025}. 

\subsection{Comparison with theoretical SNe dust attenuation laws}

We also perform our analysis using attenuation laws derived from theoretical modeling. We keep all other aspects of the data and fitting the same as discussed in Section \ref{sec:sedfits} but change the attenuation model to the extinction law from \cite{Hirashita2005}, and separately the attenuation law from \cite{Narayanan2025}, both of which are shown on Figure \ref{fig:av}. The \cite{Hirashita2005} extinction law is calculated from semi-analytic modeling of Pop III SNe \citep[c.f.,][]{Nozawa2003}, whereas \cite{Narayanan2025} perform cosmological zoom-in simulations down to $z=6$ using the dust prescriptions of \cite{Narayanan2023} which implements grain formation and evolution in the sub-grid physics.

The \cite{Hirashita2005} law is Calzetti-like at optical wavelengths with a UV slope more similar to the SMC law and close to that of \cite{Markov2024}. When fitting the data with this model (notably just extinction) we find that the posteriors on stellar mass, $M_{\rm UV}$, and $\chi^2_{r\,\rm UV}$ are comparable to when assuming SMC dust. This arises from the fact that the blue rest-frame UV colors of high-redshift \textit{JWST} sources preclude high $A_V$ for very steep attenuation laws that would otherwise redden the UV continuum and decrease the observed $\beta$. Based on $\chi^2_{r\,\rm UV}$ and the overall $\chi^2$, the data favor flatter attenuation laws than the \cite{Hirashita2005} model.

The attenuation law simulated by \cite{Narayanan2025} for $z>8$ galaxies is effectively flat throughout the UV and optical owing to an extreme deficit in the ISM content of very small dust grains. As discussed in \cite{Nozawa2007} this may arise when the small grains are either destroyed in the postshock flow or trapped in the dense shell, never mixing with the ambient ISM. This can have profound implications on inferring the presence and impact of dust. When modeling the full sample with this attenuation law we find high $A_V\in(0.5-1.5)$ and stellar masses larger than other estimates by $\sim1-2$ dex. However, it is unlikely that such flat attenuation laws can exist at redshifts $z\lesssim9-10$ by which point grain growth becomes highly efficient \citep{Narayanan2023}, with significant increases seen in the relative number of small grains. Among the $z>10$ subset of galaxies the \cite{Narayanan2025} attenuation law prefers $A_V=1\pm0.6$ on-average with stellar masses $0.4$ dex greater than what would otherwise be inferred assuming SMC-like dust attenuation, and as high as $6\times10^{9}\,M_\odot$. Such masses are expected to be rare \citep[e.g.,][]{Casey2024}, but are allowed under $\Lambda$CDM.

\section{Discussion\label{sec:discussion}}
As discussed in Section \ref{sec:sne:dustmass} changes to $A_\lambda$ at UV wavelengths are driven by very small dust grains that make up $<1\%$ of the total dust mass. Nevertheless, we make a first-order estimate of the dust masses in our sample as a consistency check against grain formation models arising from SNe. 
Direct observations of cold dust continuum along the Rayleigh-Jeans tail are the most robust method of measuring the total mass of dust \citep{Scoville2016,Scoville2017}. In place of such constraint we follow \cite{Casey2024lrds} in estimating $M_{\rm dust}$ from $A_V$ and the observed galaxy sizes. This method follows from the scaling relation between $A_V$ and the dust mass surface density from \cite{Draine2014}\footnote{$A_V=0.74\times\Sigma_{\rm dust}/10^{5}\,{\rm M_\odot\,{kpc^{-2}}\,[mag]}$}. See \cite{Casey2024lrds} for a detailed description. 
We use sizes as measured from F444W imaging to convert between the dust mass surface density and total dust mass. This method also allows us to test if the range in $A_V$, stellar mass and redshifts are plausibly consistent with dust formation mechanisms implicitly tied to our SNe attenuation assumption.

\begin{figure*}[ht!]
    \centering
    \includegraphics[width=\textwidth]{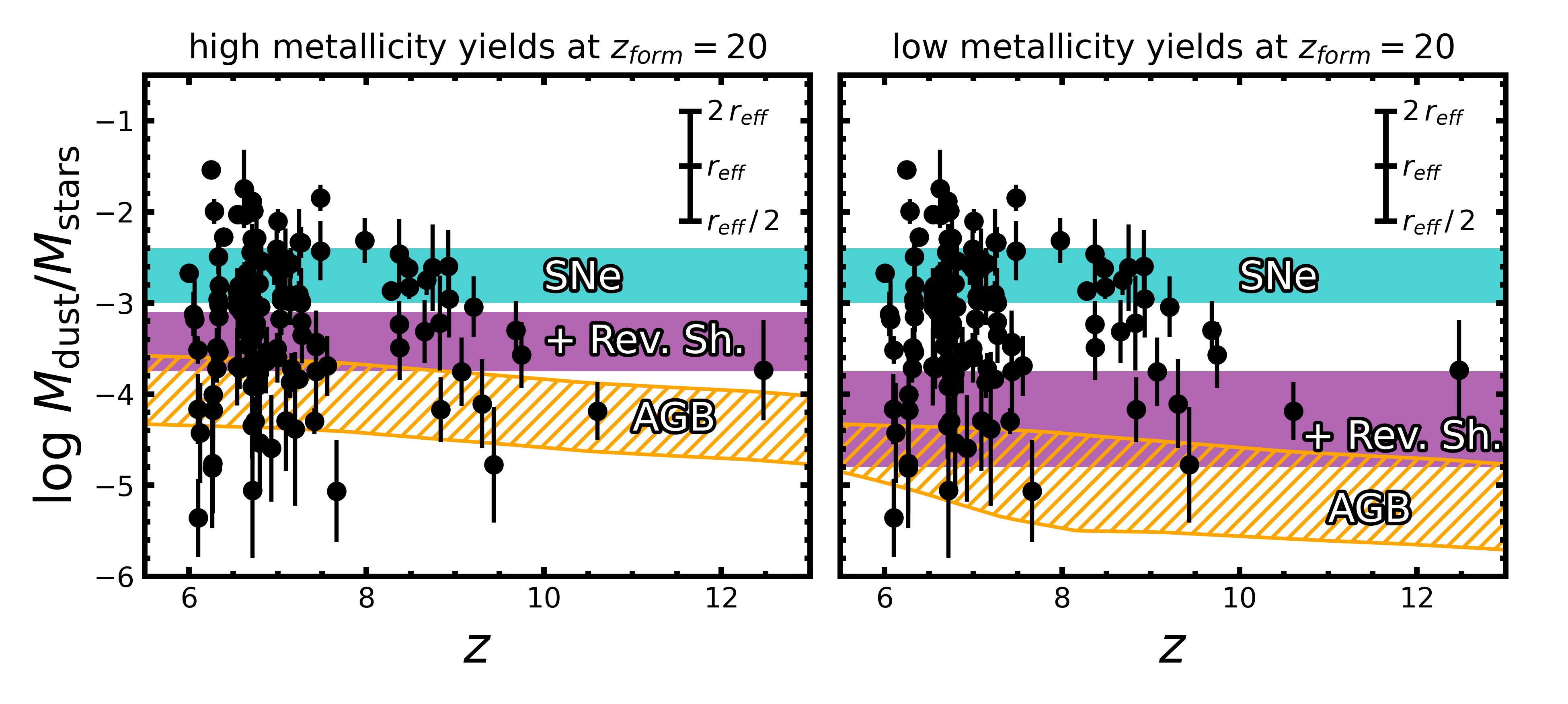}
    \caption{Total dust-to-stellar mass ratio as a function of redshift with dust masses predicted from $A_V$ following \cite{Casey2024lrds}. We overlay the models of \citet{Schneider2024} for SNe with (blue) and without (purple) reverse shock processing, as well as for AGB stars (orange hatched). The left and right panels show SNe and AGB yields when the initial metallicity at $z=20$ is above or below $0.1Z_\odot$ respectively. In the upper right corner we illustrate the change in $\log\,M_{\rm dust}/M_*$ that would arise from factors of two in radius. The predicted dust-to-stellar mass ratios are largely within the range of SNe yields with $60\%$ favoring reverse shock processing and the remaining $40\%$ requiring high yields with no destruction during the reverse shock. We emphasize that accurate dust masses are needed to validate the adoption of pure SNe dust attenuation laws; however, the dust-to-stellar mass ratios we infer from the modeled $A_V$ are to first-order self-consistent with SNe yields assuming that the dust fully covers the extent of the rest-frame optical-emitting regions, and that the dust is optically thin.} 
    \label{fig:mdms}
\end{figure*}

Figure \ref{fig:mdms} shows the resulting total dust-to-stellar mass predictions as a function of redshift. We overlay evolutionary models from SNe with and without reverse shock processing, as well as AGB stars from \citet{Schneider2024} that all begin at a formation redshift of $z=20$. We infer total dust-to-stellar mass ratios between $10^{-5}$ and $10^{-2}$. Approximately $60\%$ of the sample are consistent with SNe dust yields processed by reverse shocks. The remaining $40\%$ are consistent with no shock processing; however, assuming the dust to be twice as compact as the F444W half-light radius would bring most of these into agreement with the rest of the sample. About $40\%$ of the sample, including the two $z>10$ sources, are consistent with AGB yields but would need initial metallicities $0.1<Z/Z_\odot<1$ at $z=20$ which could be difficult to achieve. The $z>10$ galaxies exhibit dust-to-stellar mass ratios from our SNe dust models fully consistent with SNe yields processed by reverse shocks with $Z/Z_\odot<0.1$ at $z_{\rm form}=20$. Better constraints on both stellar mass and the total dust masses at these high-redshifts are needed to discriminate between the dust formation scenarios, and the effects of grain growth in the diffuse ISM may also play an important role \citep{Draine1979,Draine2009,Popping2017}. But as it stands the $A_V$-inferred dust masses coming from our SNe dust fitting are plausible given the current state of modeling dust yields at $z>6$. 

The decrease in observed ${\rm M_{UV}}$ we find when modeling dust attenuation with fairly flat attenuation, as may arise from purely SNe dust, might be important for interpreting $z>8$ galaxies within a cosmological context. For instance, \cite{Munoz2024} argue that current \textit{JWST} observations predict far too many ionizing photons that would reionize the Universe by $z\sim10$ in tension with the observed decline in neutral Hydrogen fraction between $z\sim9\rightarrow6$. Essentially every parameter thought to regulate the ionizing photon rate is a function of ${\rm M_{UV}}$, including the UV luminosity function and the production rate of ionizing photons per galaxy. Moreover, the ionizing photon escape fraction could be further impacted through the diverging properties of dust between the high-$z$ galaxies driving reionization and the low-$z$ counterparts where empirical relations are calibrated~\citep{Chisholm22}. Thus, the effects of a systematic decrease in observed ${\rm M_{UV}}$ due to SNe dust attenuation are non-linear and warrant more sophisticated modeling to better understand, which we leave to future work. But the trend is at least in the right direction in resolving the present tension by allowing for fewer ionizing photons with lower ${\rm M_{UV}}$ per galaxy. 

Likewise, many works have found an over-abundance of bright sources compared to pre-\textit{JWST} predictions \citep[e.g.,][]{Adams2023,Harikane2023,Harikane2024,Finkelstein2023,McLeod2024,Whitler2025}.
This has been interpreted as evidence for high star-formation efficiencies \citep{Dekel2023}, an absence of dust \citep{Ferrara2023,Ferrara2024lya,Ferrara2024bm,Ziparo2023,Fiore2023}, or bursty star-formation \citep{Mason2023,Munoz23,Sun23}. Our modeling of SNe dust finds lower observed $\rm{M_{UV}}$ which imply lower star-formation efficiencies; however, we also infer larger stellar masses which could further exacerbate the inconsistencies with pre-\textit{JWST} predictions. In either case, significant uncertainty remains in making reliable stellar mass measures for which understanding the dust attenuation law is just one of many important components. 

\subsection{The impact of geometry and orientation on gray attenuation}
The line of sight by which a galaxy is observed through plays a crucial role in setting both the dust observed in emission \citep{Cochrane2024} as well as the corresponding dust attenuation law \citep[e.g.,][]{Salim2020,Sommovigo2025}. The strongest variations in attenuation arise for disk systems whereby the relative inclination with respect to the observer plays a dominant role. At higher redshifts galaxies become less disky \citep{vanderWel2014,Zhang2019}, which may mitigate the complexity between the overall orientation and observed attenuation law. However, internal geometry is also important for setting the shape of attenuation. Generally, attenuation curves become grayer as the distribution of stars, gas and dust becomes less homogeneous and the stellar light is dominated by young stars \citep{Witt1996,Natale2015,Seon2016,Narayanan2018}. Thus, inhomogenous mixing of these components in young systems could preserve the intrinsically gray extinction of SNe dust. Indeed, galaxies at higher redshifts trend towards more irregular geometries as seen by high-resolution observations with \textit{JWST} and ALMA \citep{Bowler2022,HuertasCompany2024,Lines2024}, and simulations have found that attenuation tends to converge to extinction at $z\sim6$ \citep{Narayanan2018}. The relative importance of geometry on attenuation at very high redshifts could be constrained with future spatially resolved observations of Balmer decrements.


\section{Conclusion}
We use flat dust attenuation laws representative of the dust grain size distributions expected to arise purely from supernovae (SNe) to model the spectral energy distributions of $z_{\rm spec}>6$ galaxies. SNe dust has an intrinsically steeper grain size distribution characterized by fewer small ($a\sim0.01\,\mu m$) grains. This produces a much flatter UV attenuation curve than an SMC- or MW-like law. We implement a flat attenuation law, as inferred empirically by \cite{Gao2020}, in \texttt{BAGPIPES} \citep{Carnall2018}, a spectral energy distribution modeling code. We then use \texttt{BAGPIPES} and the SNe dust attenuation law\footnote{\url{https://github.com/jed-mckinney/bagpipes-sne-dust}} to model the \textit{JWST}/NIRCam photometry from JADES \citep{Eisenstein2023} for 138 $z_{\rm spec}>6$ galaxies. Our main conclusions are: 
\begin{enumerate}
    \item A fairly flat dust attenuation law lets stellar population synthesis models find better fits to the rest-frame UV observation of $z\sim6-12$ galaxies. 
    \item We recover a relation in UV slope and observed UV luminosity generally consistent with existing works that assume different dust attenuation laws. 
    \item We find systematically fainter observed UV luminosities by $0.23$ mag on-average when assuming flat UV attenuation to model SNe dust because the models can find better fits with the gray attenuation law.
    \item When assuming a fairly flat attenuation vs.~SMC-like laws, total stellar masses increase by $0.1^{+0.2}_{-0.1}$ dex on-average, and $A_V$ are higher by $0.3^{+0.2}_{-0.2}$ mag. 
    \item The choice of which attenuation law used to represent SNe dust has profound impacts on the inferred physical quantities of the galaxies. Using laws with steep rest-frame UV slopes (small grains) converge on low $A_V$ solutions because any amount of UV reddening makes it difficult for the underlying stellar population models to reach the very blue observed colors. In contrast, purely flat attenuation laws that may be possible at $z>10$ \citep[c.f.,][]{Narayanan2025} allow for $A_V\sim1$ at $z=12$ without leading to unphysical total stellar masses.
    \item To first order SNe attenuation laws yield self-consistent total dust-to-stellar mass ratios given the formation timescales needed to build these dust reservoirs rapidly in the early Universe.
\end{enumerate}
The wide-spread existence of significant dust reservoirs at $z>6$ and the origins of that dust if present remains an open question. More work is needed to observationally confirm and characterize dust at early cosmic times, which can be achieved in the UV/optical through attenuation modeling or Balmer decrements, and also by observations at far-infrared/sub-mm wavelengths to observe dust in emission. Similarly, there is a strong need for theoretical work and local observations to better constrain the yields from all grain formation channels that may or may not be relevant. The impacts of dust assumptions are as important to the stellar mass and UV luminosity functions at these redshifts as variations in the underlying stellar models and star-formation histories.

\vspace{10pt}
{\small
JM thanks the UT Austin Cosmic Frontier Center community for the engaging discussions which motivated this work, as well as Brandon Hensley and Desika Narayanan for their invaluable insight into dust attenuation and for providing access to models. 
JM acknowledges the invaluable labor of the maintenance and clerical staff at our institutions, whose contributions make our scientific discoveries a reality. JM thanks NASA and acknowledges support through the Hubble Fellowship Program, awarded by the Space Telescope Science Institute, which is operated by the Association of Universities for Research in Astronomy, Inc., for NASA, under contract NAS5-26555. OC and HA acknowledge support by the National Science Foundation Graduate Research Fellowship under grant number DGE 2137420. JBM acknowledges support from NSF Grants AST-2307354 and AST-2408637, and by the NSF-Simons AI Institute for Cosmic Origins.

Authors from UT Austin acknowledge that UT is an institution that sits on indigenous land. The Tonkawa lived in central Texas, and the Comanche and Apache moved through this area. We pay our respects to all the American Indian and Indigenous Peoples and communities who have been or have become a part of these lands and territories in Texas. We are grateful to be able to live, work, collaborate, and learn on this piece of Turtle Island. 

This work is based [in part] on observations made with the NASA/ESA/CSA James Webb Space Telescope. The data were obtained from the Mikulski Archive for Space Telescopes at the Space Telescope Science Institute, which is operated by the Association of Universities for Research in Astronomy, Inc., under NASA contract NAS 5-03127 for JWST. These observations are associated with programs GTO \#1180, \#1181, \#1210, \#1286, and GO \#1895, \#1963, and \#3215. 
}

\bibliography{references}
\bibliographystyle{aasjournal}

\end{document}